\documentclass[a4paper]{jpconf}
\usepackage{graphicx}
\usepackage{iopams}
\usepackage{dcolumn}
\usepackage{bm}
\usepackage{amsmath}
\newcommand{\LFAO}{LaFeAsO}
\newcommand{\BFA}{BaFe$_2$As$_2$}
\newcommand{\SFA}{SrFe$_2$As$_2$}
\newcommand{\MFA}{$M$Fe$_2$As$_2$}
\newcommand{\LFAOX}{LaFeAsO$_{1-x}$F$_x$}
\newcommand{\BFAX}{Ba$_{1-y}$K$_{y}$Fe$_2$As$_2$}
\newcommand{\SFAX}{Sr$_{1-y}$K$_{y}$Fe$_2$As$_2$}

\newcommand{\dxy}{\ensuremath{d_{xy}}}

\newcommand{\dxyz}{\ensuremath{d_{yz,zx}}}

\newcommand{\mb}{\ensuremath{\mu_{\text{B}}}}

\newcommand{\qv}{\ensuremath{\mathbf{q}}}

\newcommand{\tv}{\ensuremath{\mathbf{t}}}

\newcommand{\Rv}{\ensuremath{\mathbf{R}}}

\newcommand{\XB}{\ensuremath{\bar{\mathrm{X}}}}
\newcommand{\MB}{\ensuremath{\bar{\mathrm{M}}}}
\newcommand{\GB}{\ensuremath{\bar{\Gamma}}}
\newcommand{\Eq}{\ensuremath{E(\mathbf{q})}}

\begin{document}

\title{Magnetic properties of iron pnictides from spin-spiral calculations}
\author{Alexander Yaresko}

\address{Max-Planck-Institut f\"ur Festk\"orperforschung,
  Heisenbergstra{\ss}e 1, D-70569 Stuttgart, Germany}

\ead{a.yaresko@fkf.mpg.de}

\begin{abstract}
  The wave-vector $(\qv)$ and doping dependences of the magnetic energy, iron
  moment, and effective exchange interactions in \LFAO, \BFA, and \SFA\ are
  studied by self-consistent LSDA calculations for co-planar spin spirals. For
  the undoped compounds, the calculated total energy, \Eq, reaches its minimum
  at \qv\ corresponding to stripe anti-ferromagnetic (AF) order.  In \LFAO,
  this minimum becomes flat already at low levels of electron-doping and
  shifts to an incommensurate \qv\ at $\delta$=0.2, where $\delta$ is the
  number of additional electrons ($\delta>0$) or holes ($\delta<0$) per Fe. In
  \BFA\ and \SFA, stripe order remains stable for hole doping down to
  $\delta=-0.3$. Under electron doping, on the other hand, the \Eq\ minimum
  shifts to incommensurate \qv\ already at $\delta$=0.1.
\end{abstract}

Iron pnictides have attracted great interest of both experimentalist and
theoreticians after the discovery of superconductivity with $T_{c}$=27 K in
F-doped \LFAOX\ \cite{KWHH08} and with even higher $T_{c}$ of 38 K in
oxygen-free, potassium-doped \BFAX \cite{RTJ08}. Both families of iron
pnictides have a quasi two-dimensional (2D) tetragonal crystal structure, in
which FeAs layers are separated by either LaO or Ba layers. The Fe ions form a
square lattice sandwiched between two As sheets shifted so that each Fe is
surrounded by a slightly squeezed As tetrahedron.  At about 150 K, both
stoichiometric parent compounds undergo a structural transition at which the
symmetry of the lattice lowers to orthorhombic \cite{NKKH+08,RTJS+08}.  In
\BFA\ and \SFA, stripe antiferromagnetic (AF) order of 0.4--0.8 \mb Fe moments
\cite{RTJS+08,HQBG+08}, aligned ferromagnetically (FM) along the shorter
$b$-axis and antiferromagnetically along the longer $a$-axis and the $c$-axis
\cite{HQBG+08,ZRLC+08}, sets in at the same temperature as the structural
transition. Transition to a phase with the same stripe AF order occurs also in
LaFeAsO, but at a 20 K lower temperature than the structural transition
\cite{CHLL+08}.
Electron doping of the FeAs layers in \LFAOX\ suppresses the structural and
magnetic transitions in favor of superconductivity already at $x$=0.03
\cite{DZXL+08}. Also hole doping in \BFAX\ suppresses the structural and
magnetic transition \cite{RTJ08} but at much higher doping of $\sim$0.15 holes
per Fe \cite{CRQB+08}, whereas AF fluctuations are observed up to as high K
content as $y\sim$0.5. Electron doping caused by Co substitution in
Ba(Fe$_{1-x}$Co$_x)_2$As$_2$, on the other hand, suppresses the magnetic
transition at $x\sim$0.06. 

Since in many of the superconducting iron pnictides the highest $T_c$ is
observed at the doping level at which the magnetic transition is suppressed,
the superconductivity seems to be closely related to magnetism and
understanding the magnetic interactions between Fe moments in these compounds
is of utmost importance. Although the theoretically calculated Fe magnetic
moment and the stabilization energies of different magnetic solutions depend
strongly on the employed computational method and exchange-correlation
functional all band structure calculations that the stripe AF order is the
magnetic ground state in both parent compounds
\cite{YLHN+08,Yil08,MSJD08,OKZG+09}.

Results of spin-spiral calculations and an analysis of effective exchange
interactions between Fe moments in electron doped \LFAO\ and hole doped \BFA\
were presented in our previous paper \cite{YLAA09}. In this paper we show that
electron doping of FeAs layers in $M$(Fe$_{1-x}$Co$_x)_2$As$_2$ ($M$=Ba, Sr)
destabilizes stripe AF order as efficiently as in \LFAO.

Self-consistent calculations for co-planar spin spirals in \LFAO\ and \MFA,
with $M$=Ba or Sr, were performed within the LSDA using the linear muffin-tin
orbital (LMTO) method in the atomic sphere approximation \cite{And75}.
The effect of doping was simulated by using the virtual crystal approximation,
with the doping level $\delta$ defined as the deviation of the number of
valence electrons from corresponding values for undoped compounds normalized
to the number of Fe atoms in the unit cells. Positive and negative values of
$\delta$ correspond to electron and hole doping, respectively.  Calculations
for \LFAO\ were performed for $\delta$=0.1, 0.2, and 0.3, which corresponds to
F content of $x$=$\delta$ in \LFAOX. Potassium doping of $y$=0.2, 0.4, 0.6 in
\BFAX\ and \SFAX\ was modeled by $\delta$=$-$0.1, $-$0.2, $-$0.3
($y=2|\delta|$). Finally, electron doping of FeAs layers in
$M$(Fe$_{1-x}$Co$_x)_2$As$_2$ caused by Co substitution was studied for
$\delta$=0.1.

The calculations for all doping levels were carried out for experimental room
temperature crystal structures of the undoped compounds.  The tetragonal
($P4/nmm$) unit cell with $a$=4.0353 \AA, $c$=8.7409 \AA,
$z_{\text{La}}$=0.14154, and $z_{\text{As}}$=0.6512 was used for \LFAO\
\cite{KWHH08}. \MFA\ band structures were calculated for the body centered
tetragonal ($I4/mmm$) unit cell with $a$=3.9625 \AA, $c$=13.0168 \AA, and
$z_{\text{As}}$=0.3545 for $M$=Ba \cite{RTJS+08} and $a$=3.9243 \AA,
$c$=12.3644 \AA, and $z_{\text{As}}$=0.36 for $M$=Sr \cite{TRWS+08}.

As explained in details in \cite{YLAA09}, LMTO calculations for \LFAO\ place
Fe \dxy\ bands too close to \dxyz-derived bands as compared to results
obtained with the full potential linear augmented plane wave (LAPW) method
\cite{MJBK+08}. Better agreement between two band structures can be obtained
by adding in LMTO calculations an on-site shift of $-$150 meV to the Fe \dxy\
states. In the following, the results calculated for \LFAO\ with the shifted
Fe \dxy\ states are shown.

\begin{figure}[ht]
\includegraphics[width=0.45\textwidth]{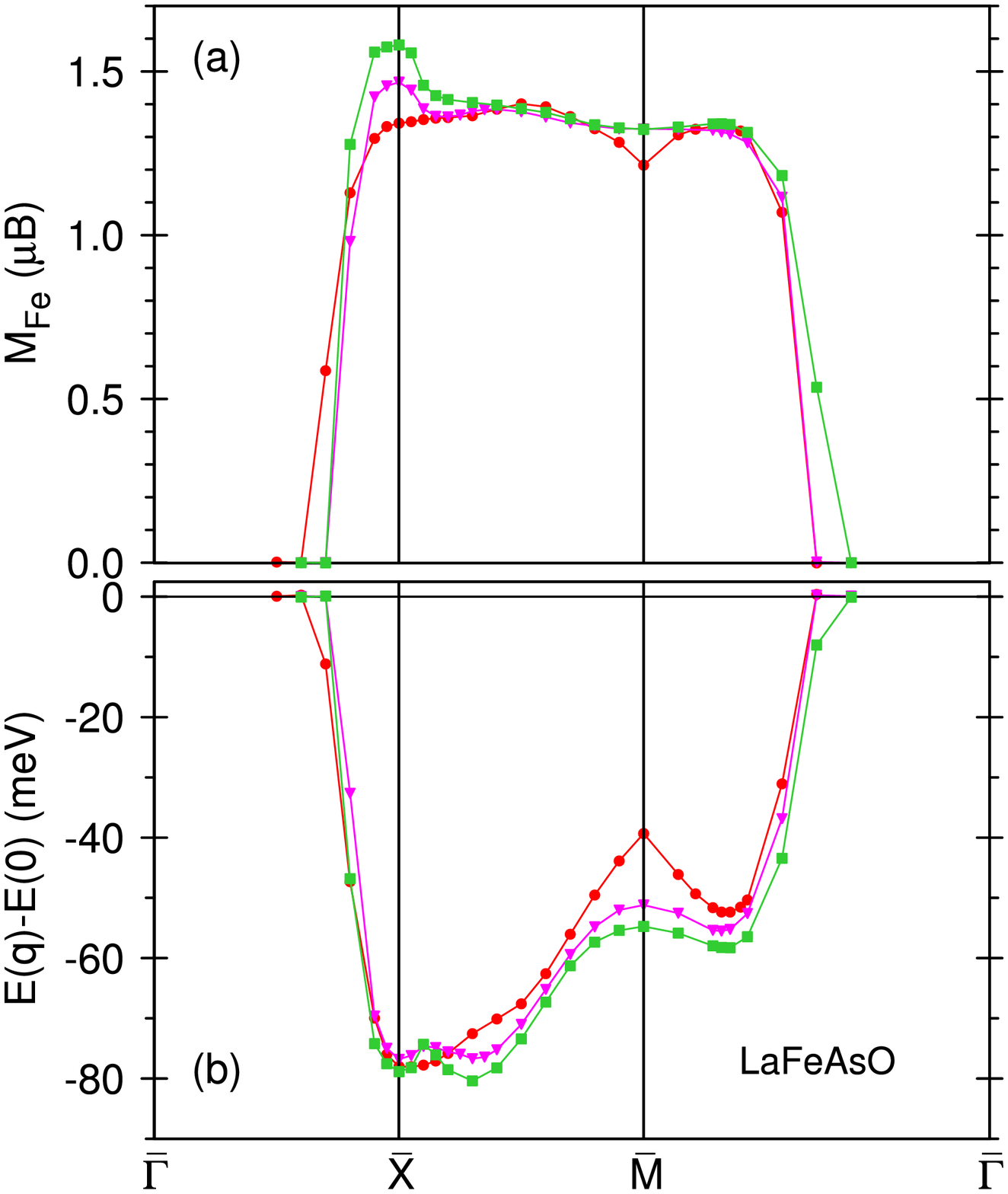}
\hspace{1em}
\begin{minipage}[b]{0.45\textwidth}
\caption{\label{fig:emq_La} The \qv\ dependencies of the Fe magnetic moment (a)
  and the total energy (b) calculated for \LFAO\ with 
  $\delta$=0 (red \fullcircle), 0.1 (magenta $\blacktriangledown$), and 0.2
  (green $\blacksquare$).}
\end{minipage}
\end{figure}

Spin spiral calculations were performed on the base of the generalized Bloch
theorem \cite{San91a} under the assumption that the direction of the
magnetization is constant within each atomic sphere. In these calculations the
angle $\theta$ between Fe magnetic moments and the $c$ axis is set to 90
degree, i.e., all Fe moments lie in the $ab$ plane. The angle $\phi$ between
the Fe moment and the $a$ axis depends on the wave vector \qv\ of a spiral
and is given by $\phi_i=\phi^{0}_{i}+\qv \cdot\Rv$, where \Rv\ is the
translation vector. The initial phases $\phi^{0}_{i}$ for two Fe sites at the
positions $\tv_i$ in the tetragonal unit cell are fixed by $\phi^{0}_{i}=\qv
\cdot\tv_i$.

The total energy of spin-spirals was calculated for \qv\ varying along \GB\
(0,0) -- \XB\ ($\pi/d$,0) -- \MB\ ($\pi/d$,$\pi/d$) -- \GB\ lines in the two
dimensional magnetic Brillouin zone (BZ), where $d$ is the distance between Fe
nearest-neighbors in the $ab$ plane. At \XB\ and \MB\ points collinear AF
stripe- and checkerboard-like spin structures are formed. The calculations
were performed for FM ($q_z$=0) and AF ($q_z=\pi/d_{\perp}$) alignments of Fe
moments in adjacent Fe layers, were $q_z$ is the $z$ component of the wave
vector and $d_{\perp}$ is the interlayer distance.

\begin{figure}[ht]
\begin{center}
\includegraphics[width=0.90\textwidth]{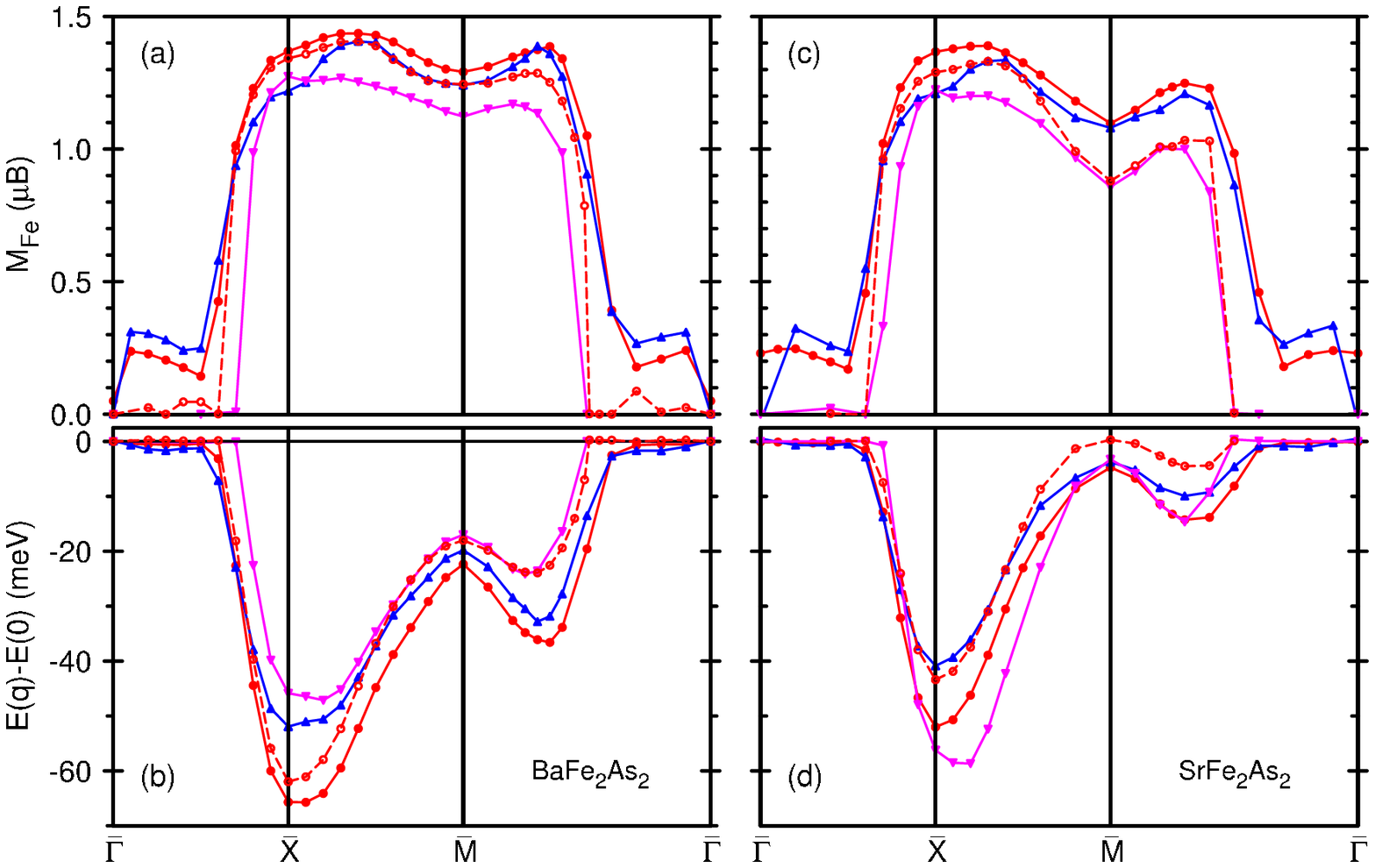}
\end{center}
\caption{\label{fig:emq_BaSr} The \qv\ dependencies of the Fe magnetic moment
  (a, c) and the total energy (b, d) calculated for \BFA\ (a, b) and \SFA\ (c,
  d) with AF alignment of Fe moments along the $c$ axis ($q_z=\pi/d_{\perp}$)
  and $\delta$=$-$0.1 (blue $\blacktriangle$), 0 (red \fullcircle), and 0.1
  (magenta $\blacktriangledown$). The corresponding curves calculated for
  undoped compounds ($\delta$=0) with $q_z$=0 are also shown (red
  \opencircle).}
\end{figure}

By comparing the total energies calculated for $q_z$=0 and $q_z=\pi/d_{\perp}$
we found that the interlayer magnetic coupling in \LFAO\ is extremely weak.
The \qv\ dependencies of the Fe magnetic moment and the total energy per Fe,
\Eq, calculated for \LFAO\ with $q_z$=0 and $\delta$=0, 0.1, and 0.2 are shown
in figure \ref{fig:emq_La}.  The minimum of \Eq\ for the undoped compound is
found at the \XB\ point, i.e., for stripe AF order. This, together with the
appearance of a local minimum along the \MB--\GB\ line, 
suggests that
effective interactions between the nearest ($j_1$) and next-nearest ($j_2$) Fe
neighbours are antiferromagnetic with $j_2>j_1/2$ \cite{YLAA09}.
The energy of spin spirals rapidly increases with the decrease of $|\qv|$ and
and at short wave vectors the self-consistent solution becomes nonmagnetic.
As a FeAs layer in \LFAO\ is doped with
electrons ($\delta>0$) to simulate F doping in \LFAOX, another local minimum
develops at an incommensurate wave vector $\qv_{min}$ along the \XB--\MB\
line. $E(\qv_{min})$ becomes lower than $E(\XB)$, i.e., the collinear
stripe-like ordered AF solution becomes unstable, at $\delta \gtrsim$0.1.

In contrast to \LFAO, the energy of spin spirals calculated for \MFA\ with AF
alignment of Fe moments along the $c$ direction is significantly lower than
for FM one. The corresponding curves for the undoped compounds are compared in
figure \ref{fig:emq_BaSr}. The change of magnetic order in Fe chains running
along $c$ from AF to FM costs at the \XB\ point 3.5 meV/Fe in \BFA\ and 8.6
meV/Fe in \SFA. The \qv\ dependencies of the total energy for both undoped
\MFA\ compounds are similar to \Eq\ for \LFAO, with the minimum at \XB\ and a
local one along the \MB--\GB\ line. Although the stabilization energy of the
stripe-ordered solution, $|E(\XB)-E(0)|$, is somewhat lower in \SFA\ than in
\BFA, the energy difference $E(\MB)-E(\XB)$ of 38 meV/Fe in the former is
larger than in the latter (29 meV/Fe).
This indicates that also $j_1$ and $j_2$ couplings within a FeAs layer are
stronger in \SFA\ than in \BFA.

As it was shown in \cite{YLAA09}, when \BFA\ is doped with holes, stripe AF
order remains stable up to the doping level of $\delta\sim -0.3$, which
corresponds to K content of about 0.6 in \BFAX. The stabilization energy of
the stripe-ordered solution is, however, strongly reduced by hole doping. The
same tendency is also obtained from spin spiral calculations for \SFA\ with
$\delta <$0, which is illustrated by $\delta = -0.1$ curves in figure
\ref{fig:emq_BaSr}.
Calculations performed for electron-doped \MFA\ show that the total energy
minimum shifts from \XB\ to incommensurate \qv\ already for
$\delta$=0.1. These results agree with the experimental phase diagram of
Ba(Fe$_{1-x}$Co$_x)_2$As$_2$ according to which the structural and magnetic
transition at $T\sim$140 K is suppressed at $x\approx 0.06$ \cite{CAKF09}.

Recently, strongly anisotropic nearest neighbour couplings along AF ($j_{1a}$)
and FM ($j_{1b}$) Fe chains have been obtained from linear response
calculations \cite{HYPS09}.  We estimated $j_{1a}$, $j_{1b}$, and $j_{2}$ from
a list-square fit to $E(\qv)$ curves calculated using the magnetic force
theorem. Preliminary results show that the anisotropy of $j_{1a}$ and $j_{1b}$
is much weaker than reported in Ref.~\cite{HYPS09}. However, the anisotropy
strongly increases if the fit is performed in a small region of \qv-space
around the \XB\ point. The results of these calculations will be published
elsewhere.

\section*{References}
\newcommand{\noopsort}[1]{} \newcommand{\printfirst}[2]{#1}
  \newcommand{\singleletter}[1]{#1} \newcommand{\switchargs}[2]{#2#1}
\providecommand{\newblock}{}


\end{document}